\documentclass[prd,aps,10pt,twocolumn,floats,nofootinbib,superscriptaddress,eqsecnum,floatfix,showpacs]{revtex4-1}

\usepackage{amssymb,amsmath}
\usepackage{epsfig}
\pdfoutput=1
\usepackage{amssymb,amsmath}
\usepackage{epsfig}
\usepackage{xcolor}
\usepackage[utf8]{inputenc}

\newcommand{\be}{\begin{equation}}
\newcommand{\ee}{\end{equation}}
\newcommand{\ba}{\begin{eqnarray}}
\newcommand{\ea}{\end{eqnarray}}
\newcommand{\beg}{\begin{gather*}}
\newcommand{\eng}{\end{gather*}}

\newcommand{\hh}{,\hspace{0.5cm}}

\newcommand{\eq}[1]{(\ref{#1})}
\newcommand{\lap}{\bigtriangleup}

\newcommand{\ins}[1]{{\mbox{\tiny #1}}}
\newcommand{\ind}[1]{{\mbox{\scriptsize #1}}}

\newcommand{\ts}[1]{{\boldsymbol{#1}}}

\def\XXint#1#2#3{{\setbox0=\hbox{$#1{#2#3}{\int}$ }
\vcenter{\hbox{$#2#3$ }}\kern-.6\wd0}}

\usepackage[makeroom]{cancel}
\usepackage[caption=false]{subfig}
\usepackage[colorlinks=true,
            citecolor=green,
            linkcolor=red,
            filecolor=cyan,
            urlcolor=magenta,
            backref=false]{hyperref}

\newcommand{\dd}{\mbox{d}}

\begin{document}

\title{The gravitational field of static $p$-branes in linearized ghost-free gravity}

\author{Jens Boos}
\email{boos@ualberta.ca}
\affiliation{Theoretical Physics Institute, University of Alberta, Edmonton, Alberta, Canada T6G 2E1}

\author{Valeri P. Frolov}
\email{vfrolov@ualberta.ca}
\affiliation{Theoretical Physics Institute, University of Alberta, Edmonton, Alberta, Canada T6G 2E1}
\affiliation{\cal{Yukawa Institute for Theoretical Physics, Kyoto University, 606-8502, Kyoto, Japan}}
\author{Andrei Zelnikov}
\email{zelnikov@ualberta.ca}
\affiliation{Theoretical Physics Institute, University of Alberta, Edmonton, Alberta, Canada T6G 2E1}

\date{\today}

\begin{abstract}
We study the gravitational field of static $p$-branes in $D$-dimensional Minkowski space in the framework of linearized ghost-free (GF) gravity. The concrete models of GF gravity we consider are parametrized by the non-local form factors $\exp(-\Box/\mu^2)$ and $\exp(\Box^2/\mu^4)$, where $\mu^{-1}$ is the scale of non-locality.
We show that the singular behavior of the gravitational field of $p$-branes in General Relativity is cured by short-range modifications introduced by the non-localities, and we derive exact expressions of the regularized gravitational fields, whose geometry can be written as a warped metric. For large distances compared to the scale of non-locality, $\mu r\rightarrow\infty$, our solutions approach those found in linearized General Relativity.
\end{abstract}

\pacs{04.20.Dw, 04.20.Jb, 04.50.Kd, 11.27.+d \hfill Alberta-Thy-02-18}

\maketitle


\section{Introduction}

Einstein's theory of General Relativity (GR) describes gravitational physics on the scale of our solar system remarkably well, and it has a well-defined Newtonian limit in the context of weak gravitational fields. However, one of the predictions of GR is the existence of spacetime singularities: spacetime loses geodesic completeness, and the curvature diverges. These problematic features in the ultraviolet (UV) regime are deemed physically unreasonable, and hence one may ask the question: what is the correct UV completion of gravity?

A fairly generic approach is to \emph{UV complete} GR by adding terms that are quadratic or of higher order in the curvature, or contain more derivatives. This is a natural procedure since these terms are generated by the effective action during quantization anyway. For an example, see e.g.~\cite{Stelle:1976gc,Stelle:1977ry}. As it turns out, the gravitational potential of a point mass becomes regular for this class of theories (see \cite{Asorey:1996hz,Modesto:2014eta} for details). Unfortunately, the propagators of these theories contain ghost modes, reflecting an inherent instability of this class of theories \cite{Stelle:1976gc,Stelle:1977ry,Biswas:2005qr,Barnaby:2007ve}. Hence, the addition of other higher derivative terms may mitigate the UV behavior but is usually accompanied by extra unphysical ghost or tachyon modes.

An interesting approach is to consider a theory with \emph{infinitely many} derivatives, which is equivalent to a non-local modification of GR. Non-local field theories were considered a long time ago (see, e.g., \cite{Efimov1967,Efimov:19,Efimov:1972wj,Efimov:18,Efimov:1976nu}); see also \cite{Hehl:2009es,Blome:2010xn} for a more recent approach. Also, they appear naturally in the context of non-commutative geometry deformation of GR \cite{Nicolini:2005vd,Spallucci:2006zj} (see the review \cite{Nicolini:2005zi} and references therein). The initial value problem in non-local theories was studied in \cite{Barnaby:2007ve,Barnaby:2010kx}.

In some cases, with a proper choice of non-local form factors, UV singularities may be avoided while no extra propagating degrees of freedom appear.
Such modifications of GR are called \emph{ghost-free (GF) gravity} (see, e.g., \cite{Tomboulis:1997gg,Modesto:2011kw,Biswas:2011ar,Modesto:2012ys,Biswas:2013kla,Biswas:2013cha,Modesto:2014lga,Tomboulis:2015gfa,Tomboulis:2015esa,Modesto:2017sdr} and references therein). These theories are usually characterized by a mass parameter $\mu$ or a length scale $\ell\sim\mu^{-1}$ at which non-localities become important. At large scales GF gravity is expected to reproduce GR. The theory may have another energy scale, $\mu_*$, where quantum fluctuations of the metric become large. We assume that $\mu\ll \mu_*$ (or, equivalently, $\ell \gg \ell_* = \mu_*^{-1}$). In other words, in the approach adopted in this paper the metric is treated as classical, that is, its quantum fluctuations are considered to be small. Therefore the length parameter $\ell$ is chosen to be larger than the Planck length, the string scale, or some other scale, depending on the fundamental quantum theory of gravity.

Recently, the entropy of black holes \cite{Conroy:2015nva}, quantum effects like one-loop renormalization \cite{Modesto:2017hzl}, and the Unruh effect \cite{Kajuri:2017jmy,Modesto:2017ycz} have been studied in GF gravity in higher dimensions. The study of GF gravity as applied to the problem of cosmological singularities can be found in \cite{Biswas:2010zk,Calcagni:2013vra}. GF gravity applications to the problem of black hole singularities were considered in \cite{Hossenfelder:2009fc,Modesto:2010uh,Zhang:2014bea,Conroy:2015wfa,Li:2015bqa,Calcagni:2017sov,Cornell:2017irh}.
It is well known that linearized GF gravity regularizes the gravitational field of point-like sources \cite{Biswas:2011ar,Modesto:2010uh,Frolov:2015bia}, and recent studies indicate that this may remain true even in the full non-linear GF theory \cite{Koshelev:2018hpt}. Moreover, all four-dimensional curvature invariants are finite everywhere, and at the location of the point particle spacetime approaches conformal flatness \cite{Buoninfante:2018xiw}. One can expect that for black holes of large mass and size the GF modification of gravity results only in small corrections. In particular, the properties of their horizons remain practically the same. For black holes of small masses ($\sim\mu$), GF gravity effects may become very important. For example, dynamical solutions, like collapsing matter or head-on collision of particles in four and higher dimensions, were studied in \cite{Frolov:2015bta,Frolov:2015bia}, showing that there exists a mass gap for black hole formation in GF gravity.

One can expect that these properties (regularity of solutions and mass gap effect) might be valid in the complete GF gravity even in the strong field regime. Before studying this rather complicated problem it is instructive to demonstrate that linearized GR gravity regularizes the gravitational fields of not only pointlike particles, but also the field of other infinitely thin sources, e.g., cosmic strings, membranes, and $p$-branes in arbitrary dimensions. The aim of this paper is to demonstrate this explicitly and to obtain explicit solutions of the linearized GF gravity equations for such objects.

Let us first discuss the well-known example of a cosmic string \cite{Vilenkin:1981zs} in the context of GR. Expressed in Cartesian coordinates, a straight cosmic string in Minkowski space located along the $z$-axis has the stress-energy tensor
\begin{align}
T{}_{\mu\nu} = \epsilon \left( \delta{}^t_\mu \delta{}^t_\nu - \delta{}^z_\mu \delta{}^z_\nu \right) \delta(x)\delta(y) \, .
\end{align}
Defining $g{}_{\mu\nu} = \eta{}_{\mu\nu} + h{}_{\mu\nu}$, a test-string solution in the framework of linearized GR in the harmonic gauge is given by
\begin{align}
\label{eq:vilenkin-string}
h{}_{xx} = h{}_{yy} = -8 G \epsilon\ln(r/r_0) \, ,
\end{align}
where $r^2 = x^2 + y^2$ and $r_0$ is an integration constant.

To linear order the curvature tensor vanishes everywhere except for the origin of the $xy$-plane along the $z$-axis, where it has a $\delta$-like singularity producing an angle deficit. Accordingly, the geometry of a straight cosmic string corresponds to Minkowski space with a conical deficit angle $\delta = 8\pi G\epsilon$.

Is this curvature singularity along the $z$-axis cured in GF gravity? In order to answer this and similar questions in quite some generality, in this work we  study the static gravitational field of $p$-branes in $D$-dimensional Minkowski space, employing the framework of linearized GF gravity. In particular, we  show that in the case of $p=1$ and $D=4$ (cosmic strings) the conical singularity is resolved. Moreover, we  derive explicitly the regular potentials for any $p$-brane in any number of spacetime dimensions.

This paper is organized as follows: In Sec.~\ref{sec:linearization} we briefly sketch the derivation of linearized GF gravity, before discussing the general gravitational field of $p$-branes in Sec.~\ref{sec:p-branes}. In Sec.~\ref{sec:examples} we list a set of explicit examples and discuss their behavior in detail, before discussing the obtained results in Sec.~\ref{sec:conclusions}.

\section{Linearized ghost-free gravity}
\label{sec:linearization}
Consider a weak perturbation on Minkowski space in Cartesian coordinates,
\begin{align}
g_{\mu\nu}=\eta_{\mu\nu}+h_{\mu\nu} \, ,
\end{align}
where $\eta{}_{\mu\nu}$ is the Minkowski metric, and $h_{\mu\nu}$ is assumed to be small. The dynamics of $h{}_{\mu\nu}$ can be derived by variation of the GF gravity action. In order to study the action of linearized GF gravity, it is sufficient to consider only terms linear and quadratic in curvature, with generic non-local form factors. Using the symmetry properties of the Riemann tensor, the Bianchi identities, and the commutativity of the covariant derivations up to linear in the curvature terms one can show \cite{Barvinsky:1990up,Modesto:2014eta} that there are only two independent non-local form factors that characterize non-local linearized gravity in arbitrary dimensions. In $D$ dimensions, a generic action for linearized GF gravity written in Cartesian coordinates then takes the form \cite{Conroy:2015nva,Frolov:2015usa}
\begin{align}
\label{S}
S=&{1\over 2\kappa}\int \dd^D x \, \Big(
{1\over 2}h^{\mu\nu}\,a(\Box)\Box\,
h_{\mu\nu}-h^{\mu\nu}\,a(\Box)\partial_{\mu}\partial_{\alpha}\,h^{\alpha}{}_{\nu} \nonumber \\
&\hspace{45pt} +h^{
\mu\nu }\, c(\Box)
\partial_{\mu}\partial_{\nu} h
-{1\over 2}h\,c(\Box)\Box h \\
&\hspace{45pt}+{1\over
2}h^{\mu\nu}\,{a(\Box)-c(\Box)\over\Box}\partial_{\mu}\partial_{\nu}\partial_{\alpha}\partial_{
\beta}\,h^{\alpha\beta}
\Big) \, . \nonumber
\end{align}
where $a(\Box)$ and $c(\Box)$ are arbitrary, dimensionless form factors and $\kappa=8\pi G_\ins{D}$ is the gravitational constant in $D$ dimensions. This general action can be employed to describe various linearized gravitational theories \cite{Biswas:2013kla}:
\begin{itemize}
\item GR is recovered for $a=c=1$,
\item L(R) gravity for the choice $a=1$, $c=1-L''(\Box)$,
\item Weyl gravity for $L=R-\mu^{-2}C_{\mu\nu\alpha\beta}C^{\mu\nu\alpha\beta}$, $a=1-\mu^{-2}\Box$, $c=1-\mu^{-2}\Box/3$, where $\mu$ is a parameter of dimension mass.
\end{itemize}
At any rate, in order to recover GR in the infrared (IR) domain the form factors functions must satisfy the condition $a(0)=c(0)=1$.\footnote{Because the form factors $a(\Box)$ and $c(\Box)$ are dimensionless, the d'Alembertian $\Box$ can only enter these in combination with at least one length scale $\ell\sim\mu^{-1}$ via the dimensionless expression $\ell^2\Box$. Here, $\ell$ encodes the scale of non-locality and we understand the expressions $a(0)$ and $c(0)$ as the limit of $\ell \rightarrow 0$.}

Let now $\tau^{\mu\nu}$ be a stress-energy tensor of matter,
\be
\tau^{\mu\nu}={2\over\sqrt{-g}}{\delta S_\ins{matter}\over\delta g_{\mu\nu}},
\ee
then the linearized field equations for $h{}_{\mu\nu}$ become
\be\begin{split}\label{EQ1}
&\hspace{11pt}a(\Box)\big[\Box\,h_{\mu\nu}-
\partial_{\sigma}\big(\partial_{\nu}\,h_{\mu}{}^{\sigma} +\partial_{\mu}h_{\nu}{}^{\sigma}\big)\big]\\
&+c(\Box)\big[\eta_{\mu\nu}\big(\partial_{\rho}\partial_{\sigma}h^{\rho\sigma}-\Box h\big)+\partial_{\mu}\partial_{\nu}h
\big]\\
&+{a(\Box)-c(\Box)\over\Box}\partial_{\mu}\partial_{\nu}\partial_{\rho}\partial_{\sigma}h^{\rho\sigma}=-2\kappa\tau_{\mu\nu} \, .
\end{split}\ee
The resulting non-local theory of gravity is ghost-free if it is described by form factors $a(\Box)$ and $c(\Box)$ that are \emph{entire functions} of the operator $\Box$. Such entire functions can be written as the exponential of any finite polynomial. According to the classification adopted in \cite{Frolov:2015usa}, we call the simplest example of a GF theory $\mathrm{GF_1}$ with the choice
\be
a(\Box)=c(\Box)=\exp\big(-\Box/\mu^2\big) \, .
\ee
This version of GF gravity has been extensively studied in the literature \cite{Tomboulis:1997gg, Biswas:2011ar,Modesto:2011kw,Modesto:2012ys,Biswas:2013cha,Biswas:2013kla, Modesto:2014lga,Tomboulis:2015gfa,Tomboulis:2015esa,Modesto:2017sdr}, and this choice corresponds to the absence of propagating spin zero gravitational modes. The mass parameter $\mu$ describes the length scale $\ell=\mu^{-1}$ below which non-localities start become relevant and GF gravity solutions deviate significantly from those obtained in GR. Recently \cite{Frolov:2015usa}, more general theories of the type $\mathrm{GF_\ins{N}}$ with
\be
a(\Box)=c(\Box)=\exp\big((-\Box/\mu^2)^N\big)
\ee
have been studied. It was demonstrated that in the presence of time-dependent sources $\mathrm{GF_\ins{N}}$ gravities for odd $N$ suffer from instabilities, whereas they are stable in the case of even $N$.


\section{P-branes in D-dimensional Minkowski space}
\label{sec:p-branes}

\subsection{$p$-brane ansatz}

Let us consider a plane $(p+1)$-dimensional timelike surface (``$p$-brane'') embedded in $D$-dimensional Minkowski space. We use Cartesian coordinates $x{}^\mu=(t,z_a,y_i)$ such that $(t,z_a)$ are the coordinates on the $p$-brane ($a=1,\dots,p$), while $y_i$ are the spatial coordinates in the bulk directions (with $i=1,\dots,m$). Thus we have $D=1+p+m$ and the $p$-brane is located at $y_i=0$. From now on, we call the number $m$ the \emph{co-dimensionality} of the brane.

The stress-energy tensor of the brane is
\begin{align}
\label{eq:stress-energy}
T{}_{\mu\nu} = \epsilon \left( \delta{}_\mu^t \delta{}_\nu^t - \sum\limits_{a=1}^p \delta{}_\mu^{z_a} \delta{}_\nu^{z_a}\right) \, \prod\limits_{i=1}^m \delta(y_i) \, ,
\end{align}
where $\epsilon$ is the (positive) surface tension. In this approximation for the stress-energy tensor we assume that the thickness of the matter source is much smaller than the characteristic length parameter $\ell$.

The presence of stress-energy \eqref{eq:stress-energy} will lead to a deviation from Minkowski spacetime which we shall refer to as $h{}_{\mu\nu}$. As an ansatz for the perturbed spacetime we choose a warped geometry \cite{Dabholkar:1990yf,Stelle:1998xg}
\begin{align}
\dd s^2 = f(y_i) \dd \sigma^2(t, z_a) + \dd \gamma^2(y_i) \, .
\end{align}
We shall assume that the above metric in Cartesian coordinates deviates only slightly from Minkowski space. As such, there exists a gauge where $h{}_{\mu\nu}$ has the form
\begin{align}
\begin{split}
\label{eq:ansatz}
\dd s^2 &= \hspace{11pt}(1 + u)\left( -\dd t^2 + \dd z_1^2 + \dots + \dd z_p^2 \right) \\
&\hspace{12pt} + (1 + v) \left( \dd y_1^2 + \dots + \dd y_m^2 \right) \, .
\end{split}
\end{align}
Here, the functions $u$ and $v$ depend on the distance from the brane $r$ defined as $r^2 \equiv \sum_{i=1}^m (y_i)^2$. We consider these functions as perturbations, such that  $|u(r)| \ll 1 $ and $|v(r)| \ll 1$.

This geometry has the following isometries:
\begin{itemize}
\item Poincar\'e symmetry $P(1,p)$ in the $(t,z_a)$-sector, \\[-1.3\baselineskip]
\item $O(m)$ rotation symmetry in the $y_i$-sector.
\end{itemize}
The full isometry group of \eqref{eq:ansatz} is hence $P(1,p)\times O(m)$. The discrete symmetries $\mathbb{Z}_i : y_i \rightarrow -y_i$ further guarantee that the surface described by $y_i=0$ (the $p$-brane) is geodesic and hence minimal.

Let us remark that the Newtonian limit can be read off from the $h_{tt}$ component, and we define the Newtonian potential to be given by
\begin{align}
\Phi_\ind{N} := -\frac12 h_{tt} = \frac 12 u
\end{align}
for later convenience. Incidentally, for the cosmic string in GR, Eq.~\eqref{eq:vilenkin-string}, one has $\Phi_\ind{N} = 0$, implying that test particles do not feel any forces acting on them in the Newtonian limit.

\subsection{GF linearized equations and their solution}
For the warped geometry \eqref{eq:ansatz} the linearized equations \eqref{EQ1} take the following form:
\begin{align}
\label{eq:eom-inhomogeneous}
&\big[ (p+1)c(\lap) - a(\lap) \big] \lap u + (m-1)c(\lap) \lap v \\
&\hspace{34pt}= -2\kappa\epsilon \prod\limits_{i=1}^m \delta(y_i) \, , \nonumber \\
\label{eq:eom-homogeneous}
&\big[a(\lap) - (m-1)c(\lap)\big]\left( \delta_{ij} \lap - \partial_i \partial_j \right) v \\
&\hspace{34pt}- (p+1)c(\lap)\left( \delta_{ij} \lap - \partial_i \partial_j \right)u = 0 \, , \nonumber
\end{align}
where we defined $\partial_i \equiv \partial/\partial y_i$ such that $\lap := \sum_{i=1}^m \partial_i^2$ is the $m$-dimensional Laplacian, and $\delta_{ij}$ is the Euclidean metric in the $y_i$-sector. In a simple case when
\begin{align}
c(\lap) = (1 + \alpha) a(\lap) \, , \quad \alpha \not= -1 \, ,
\end{align}
the homogeneous equation \eqref{eq:eom-homogeneous} can be solved by setting
\begin{align}
\label{eq:u-v-relation}
u = \frac{1 - (m-1)(1+\alpha)}{(1+\alpha)(1+p)} v \, .
\end{align}
The remaining equation becomes
\begin{align}
\label{eq:eom-final}
f_{mp} a(\lap) \lap v = -2\kappa\epsilon \prod\limits_{i=1}^m \delta(y_i) \, ,
\end{align}
where we introduced the constant prefactor
\begin{align}
f_{mp} \equiv \frac{(1+\alpha)(m+p) - 1}{(1+\alpha)(1+p)} \, .
\end{align}
Equation \eqref{eq:eom-final} can be solved by using the method of Green functions \cite{Frolov:2015usa,Frolov:2016xhq}. Given the function $a=a(\lap)$, the Green function $D_m(r)$ for $m \ge 3$ is
\begin{align}
\label{eq:green-function}
D_m(r) = \frac{1}{(2\pi)^{\frac m2} r^{m-2}} \int\limits_0^\infty \dd \zeta \frac{\zeta^{\frac{m-4}{2}}}{a(-\zeta^2/r^2)} J_{\frac m2 - 1}(\zeta) \, ,
\end{align}
where $J_n$ denotes the Bessel function of the first kind. The Green functions for $m < 3$ can be determined via
\begin{align}
\label{eq:generate-green-functions}
\begin{split}
D_m(r) &= -2\pi\int \dd \tilde{r} D_{m+2}(\tilde{r}) \tilde{r} \\
\Leftrightarrow \quad D_{m+2}(r) &= -\frac{1}{2\pi r} \frac{\partial D_m(r)}{\partial r} \, .
\end{split}
\end{align}
The above equality follows directly from the differential properties of $J_n$ \cite{Frolov:2015usa}. The exact solution for a $p$-brane in $(D=1+p+m)$-dimensional spacetime is then given by
\begin{align}
\label{eq:general-result}
v(r) = \frac{2\kappa\epsilon}{f_{mp}} D_m(r) \, .
\end{align}
Let us notice that the radial dependence of $v$ is universal and does not depend on the dimensionality of the brane: it only depends on the brane co-dimensionality $m$. The parameter $p$ enters only in a constant prefactor. In this linear approximation to the full GF gravity, one can also consider a superposition of $\delta$-like $p$-branes: the obtained Green functions can be used to generate linearized GF gravity solutions corresponding to thick $p$-branes. It is clear that such a ``smeared'' solution remains regular and has the correct GR behavior at far distances. In a special case when the thickness of the $p$-brane is of the order of Planck length, the gravitational field at the brane position would be regular, but slightly different from the GF solution. However, at the scale $\ell\gg \ell_*$ it will coincide with the latter.

\section{$\mathrm{GF_1}$ and $\mathrm{GF_2}$ theory}
\label{sec:examples}
Let us consider the potential \eqref{eq:general-result} in the cases of $\mathrm{GF_1}$ and $\mathrm{GF_2}$ theory. That is, from now on we put $\alpha=0$ as required by $a(\Box) = c(\Box)$. In this case, the relation \eqref{eq:u-v-relation} simplifies and takes the form
\begin{align}
\label{eq:u-v-relation-simplified}
u = \frac{2 - m}{1+p} v \, .
\end{align}
One of the immediate consequences of this relation is that for co-dimension 2, that is, when $m = 2$, the  Newtonian potential $\Phi_\ind{N}$ vanishes. This is a generalization of the result for the cosmic string.

In what follows, we will first derive the explicit Green functions for $\mathrm{GF_1}$ and $\mathrm{GF_2}$ theory, and then discuss the relation to the Green functions of linearized GR.

\subsection{Green functions of $\mathrm{GF_1}$ and $\mathrm{GF_2}$ theory}
In order to distinguish the Green functions $D_m$ for different versions $\mathrm{GF}_N$ of GF gravity, we use the corresponding index $N$ as a superscript and write the Green function in the form $D^{(N)}_m$. The Green function for $m \ge 3$ in $\mathrm{GF_1}$ theory is \cite{Frolov:2015usa}
\begin{align}
\begin{split}
D_m^{(1)}(r) &= \frac{\gamma\left(\frac m2 - 1, \tfrac{\mu^2 r^2}{4}\right)}{4\pi^{m/2}r^{m-2}} \, , \\
\gamma(s,x) &:= \int\limits_0^x z^{s-1} e^{-z} \dd z \, ,
\end{split}
\end{align}
where $\gamma(s,x)$ denotes the lower incomplete gamma function. For $m=1,2,3,4$ we find the following expressions:
\begin{align}
\begin{split}
\label{eq:gf1-green-functions}
D_1^{(1)}(r) &= -\frac r2 \text{erf}\left(\frac{\mu r}{2}\right) - \frac{\exp{\left(-\mu^2 r^2/4\right)} - 1}{\sqrt{\pi}\mu} \, , \\
D_2^{(1)}(r) &= - \frac{1}{4\pi} \text{Ein}\left( \frac{\mu^2 r^2}{4} \right) \, , \\
D_3^{(1)}(r) &= \frac{\text{erf}(\mu r/2)}{4\pi r} \, , \\
D_4^{(1)}(r) &= \frac{1 - \exp(-\mu^2r^2/4)}{4\pi^2 r^2} \, ,
\end{split}
\end{align}
where Ein(x) is the complementary exponential integral and erf(x) is the error function:
\begin{align}
\begin{split}
\text{Ein}(x) &:= \int\limits_0^x \frac{1 - e^{-z}}{z} \dd z = E_1(x) + \ln x + \gamma \, , \\
E_1(x) &:= e^{-x} \! \int\limits_0^\infty \frac{e^{-z}}{z+x} \dd z \, , \quad \text{erf}(x) := \frac{2}{\sqrt{\pi}} \int\limits_{0}^x e^{-z^2} \dd z \, .
\end{split}
\end{align}
In the above, $E_1$ denotes the exponential integral, and $\gamma = 0.577\dots$ is the Euler--Mascheroni constant \cite{Olver:2010}. The expressions for $m=1,2$ have been calculated by using Eq.~\eqref{eq:generate-green-functions}, and the above set \eqref{eq:gf1-green-functions} is sufficient to calculate $D^1_m(r)$ for any $m$.

\begin{figure*}[!hbt]
  \centering
  \includegraphics[width=\textwidth]{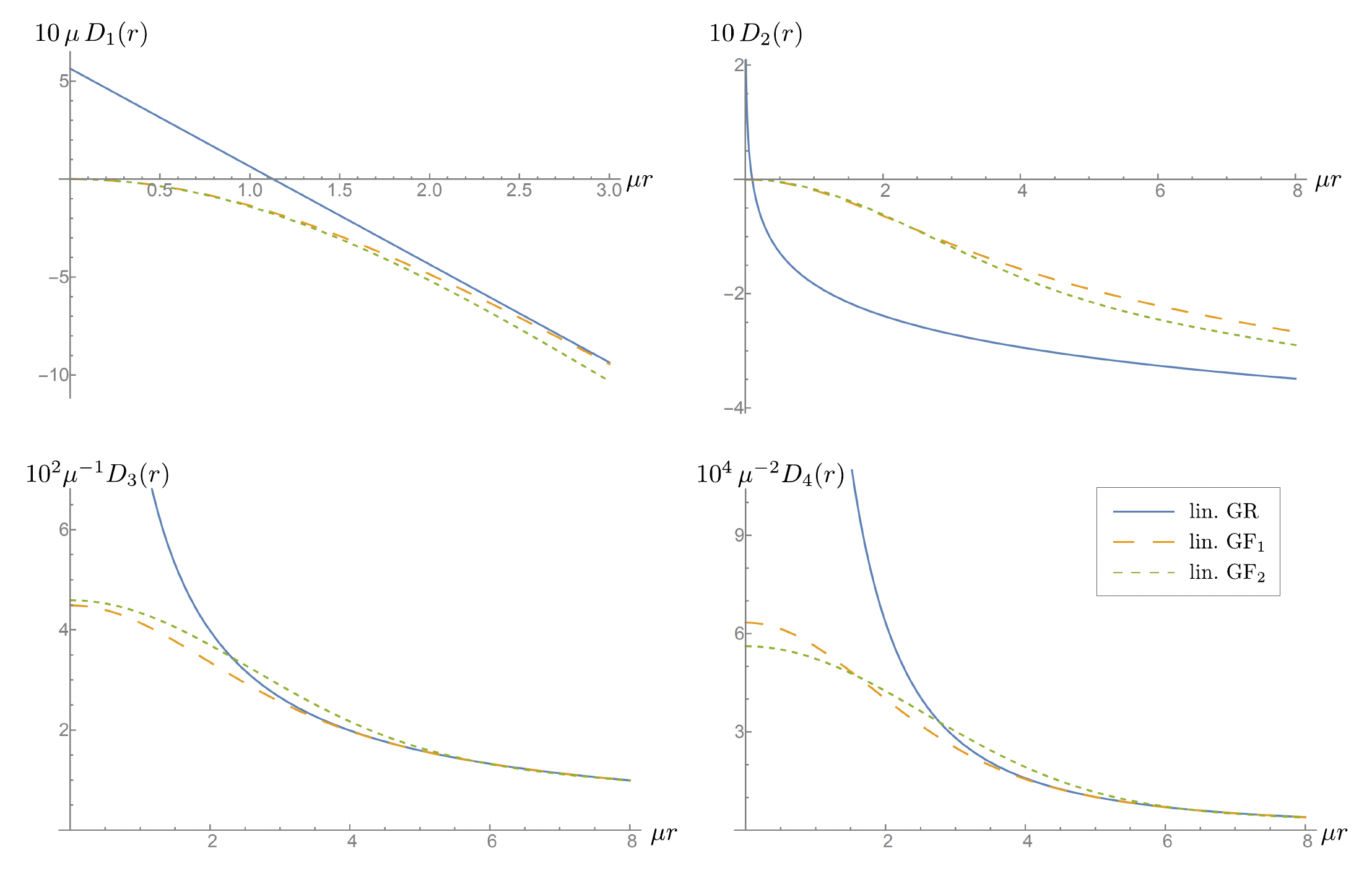}
  \caption{Dimensionless Green functions $\mu^{2-m}\,D_m(r)$ for linearized GR, $\mathrm{GF_1}$, and $\mathrm{GF_2}$ theory in the cases $m=1,2,3,4$. The solid line corresponds to linearized GR, the dashed line corresponds to linearized $\mathrm{GF_1}$ theory, and the dotted line represents linearized $\mathrm{GF_2}$ theory. All three theories agree for larges values of $\mu r$, that is, when the distance $r$ is larger than the scale of non-locality $\mu^{-1}$. For short distances, $\mu r \ll 1$, there are significant deviations between GR (which is singular at the origin) and $\mathrm{GF_1}$ and $\mathrm{GF_2}$ (which, in turn, are perfectly well-behaved).}
  \label{fig:visualization-gf1-gf2}
\end{figure*}

In $\mathrm{GF_2}$ theory, the general Green function for $m \ge 3$ takes the more complicated form \cite{Frolov:2015usa}
\begin{align}
D_m^{(2)}(r) &= \frac{\mu^{m-2}}{m(m-2)2^{\tfrac{3m}{2} - 2}\pi^{\tfrac{m-1}{2}}} \\
&\hspace{11pt}\times \Big[ \frac{m}{\Gamma\left(\frac m4\right)} {}_1\!F\!{}_3\left( \tfrac m4 - \tfrac 12;~ \tfrac 12, \tfrac m4 , \tfrac m4 + \tfrac 12;~ y^2 \right) \nonumber \\
&\hspace{11pt}-\frac{2(m-2) y}{\Gamma\left(\tfrac m4 + \tfrac 12 \right)} {}_1\!F\!{}_3\left( \tfrac m4;~ \tfrac 32, \tfrac m4 + 1, \tfrac m4 + \tfrac 12;~ y^2 \right) \Big] \, , \nonumber
\end{align}
where we defined $y \equiv (\mu^2 r^2)/16$, $\Gamma$ denotes the gamma function, and ${}_pF\!{}_q$ corresponds to the generalized hypergeometric function \cite{Olver:2010}. Again, the cases $m=1,2$ can be obtained by making use of Eq.~\eqref{eq:generate-green-functions}. We find the following expressions:
\begin{align}
\label{eq:gf2-green-functions}
D_1^{(2)}(r) &= -\frac{1}{\mu\pi} \Big\{ \hspace{6pt} 2\Gamma(\tfrac14) y\, {}_1\!F\!{}_3\left( \tfrac14;~ \tfrac34,\tfrac54,\tfrac32;~ y^2 \right) \nonumber \\
&\hspace{40pt} +\Gamma(\tfrac34) \Big[ {}_1\!F\!{}_3\left( -\tfrac14;~ \tfrac14,\tfrac12,\tfrac34;~ y^2 \right) - 1 \Big] \Big\} \, , \nonumber \\
D_2^{(2)}(r) &= -\frac{y}{2\pi} \Big[ \hspace{4pt} \sqrt{\pi}\, {}_1\!F\!{}_3\left(\tfrac12;~1,\tfrac32,\tfrac32;~y^2\right) \nonumber \\
&\hspace{40pt} - y\, {}_2 \!F\!{}_4\left(1,1;~\tfrac32,\tfrac32,2,2;~ y^2 \right) \Big] \, , \\
D_3^{(2)}(r) &= \frac{\mu}{6\pi^2}\Big[ \hspace{6pt} 3 \Gamma\!\left(\tfrac54\right) {}_1\!F\!{}_3\left( \tfrac14;~ \tfrac12,\tfrac34,\tfrac54;~ y^2 \right) \nonumber \\
&\hspace{40pt} -2y\Gamma\!\left(\tfrac34\right) {}_1\!F\!{}_3\left( \tfrac34;~ \tfrac54, \tfrac32, \tfrac74;~ y^2 \right) \Big] \, , \nonumber \\
D_4^{(2)}(r) &= \frac{\mu^2}{64\pi^2 y}\Big[ 1 - {}_0\!F\!{}_2\left( \tfrac12,\tfrac12;~ y^2 \right) + 2\sqrt{\pi} y\, {}_0\!F\!{}_2\left( 1, \tfrac32;~ y^2 \right) \Big] \, , \nonumber
\end{align}
where again $y \equiv (\mu^2 r^2)/16$.

Lastly, the Green functions of linearized GR can be obtained from both the $\mathrm{GF_1}$ Green functions $D^{(1)}_m(r)$, or the $\mathrm{GF_2}$ Green functions $D^{(2)}_m(r)$, by considering the limit $\mu \rightarrow \infty$. Denoting the GR Green functions as $D^\ind{GR}_m(r)$, we find the expressions
\begin{align}
\begin{split}
D^\ind{GR}_1(r) &= -\frac{r}{2} \, , \\
D^\ind{GR}_2(r) &= -\frac{1}{4\pi}\text{log}(r) + \gamma + \mathcal{O}(\log \mu) \, , \\
D^\ind{GR}_3(r) &= \frac{1}{4\pi r} \, , \\
D^\ind{GR}_4(r) &= \frac{1}{4\pi^2 r^2} \, .
\end{split}
\end{align}
For $D^\ind{GR}_2(r)$, the logarithmic dependence on $\log\mu$ leads to a divergence; however, it is of no physical relevance for matter sources with compact support.\footnote{If $G(x,x')$ is the Green function for the differential equation $\mathcal{D}\phi = \sigma$, then the solution is given by $\phi = \int\dd x' G(x,x')\sigma(x')$. In case the source $\sigma$ has a compact support, one can always add a constant $c$ such that $G \rightarrow G + c$. This shifted Green function still yields the same solution: $\mathcal{D} \phi = \int \mathcal{D} G(x,x') \sigma(x') + \int c \mathcal{D} \sigma = \sigma + c\int\mathcal{D}\sigma = \sigma$, where the last equality holds via integration by parts and the fact that $\sigma$ vanishes asymptotically.}

Note that both the Green functions for $\mathrm{GF_1}$ theory \eqref{eq:gf1-green-functions} and $\mathrm{GF_2}$ theory \eqref{eq:gf2-green-functions} are manifestly regular at $r=0$. In the cases $m=3,4$ their normalization has been chosen such that $D^1_m(r) = D^2_m(r) = 0$ for $r \rightarrow \infty$; in the cases $m=1,2$ their Green functions are unbounded for large radii, which is why we chose instead $D^1_m(r) = D^2_m(r) = 0$ for $r\rightarrow 0$. In contrast, the GR Green functions are singular at $r=0$.

See the plots of these Green functions in Fig.~\ref{fig:visualization-gf1-gf2}.

\subsection{Linearized curvature}
\label{sec:lin-curv}
It is also useful to study the curvature of the $p$-brane solutions. In linear approximation, the Riemann tensor, the Ricci tensor, and the Ricci scalar are \cite{Conroy:2017uds}
\begin{align}
R{}^\mu{}_{\nu\rho\sigma} &= \partial_\nu \partial{}_{[\rho} h{}_{\sigma]}{}^\mu - \partial{}^\mu \partial{}_{[\rho} h{}_{\sigma]\nu} \, \label{eq:curv-riem} \\
R{}_{\mu\nu} &= \partial{}_\alpha \partial{}_{(\mu} h{}_{\nu)}{}^\alpha - \frac 12 \left( \partial{}_\mu \partial{}_\nu h + \Box h{}_{\mu\nu} \right) \, , \label{eq:curv-ric} \\
R &= \partial{}_\alpha \partial{}_\beta h{}^{\alpha\beta} - \Box h \, . \label{eq:curv-rscalar}
\end{align}
Besides the curvature tensors and invariants, it is interesting to consider the quantity
\begin{align}
\rho_\ins{GR} &\equiv \frac{1}{\kappa} G{}_{\mu\nu}\xi{}^\mu \xi{}^\nu \, ,
\end{align}
where $G{}_{\mu\nu}$ is the linearized Einstein tensor and $\ts{\xi} = \partial_t$ is the timelike Killing vector. We see that in GR, $\rho_\ind{GR}$ corresponds to the energy density perceived by a static  observer tangential to $\ts{\xi}$.

\begin{figure*}[!hbt]
    \centering
    \subfloat{{ \includegraphics[width=0.45\textwidth]{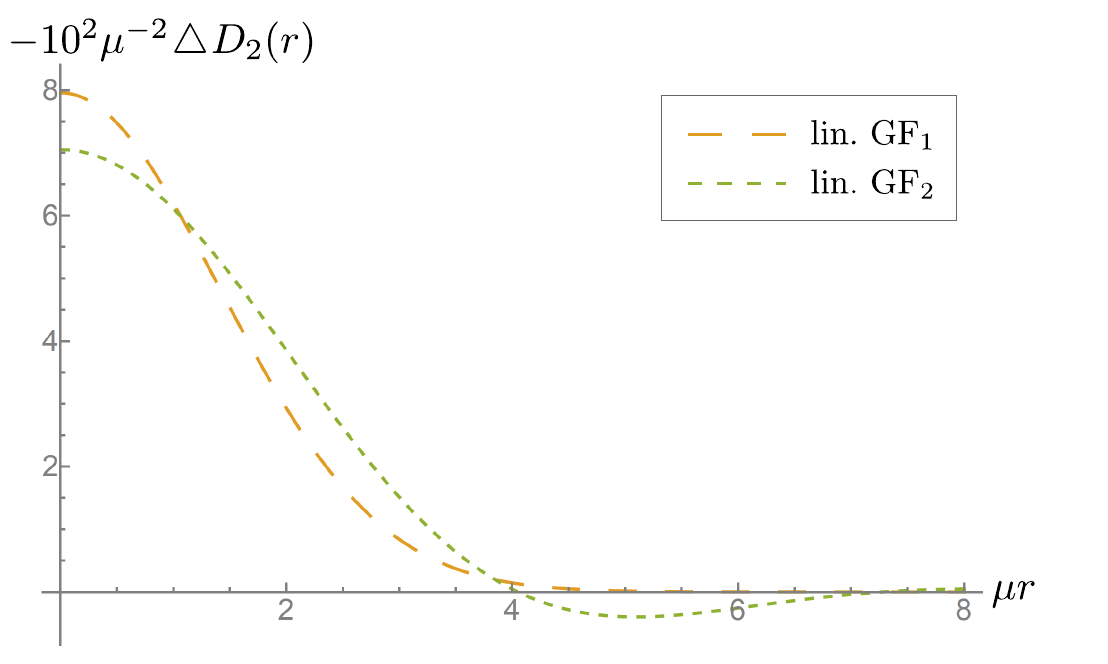} }}%
    \qquad
    \subfloat{{ \includegraphics[width=0.45\textwidth]{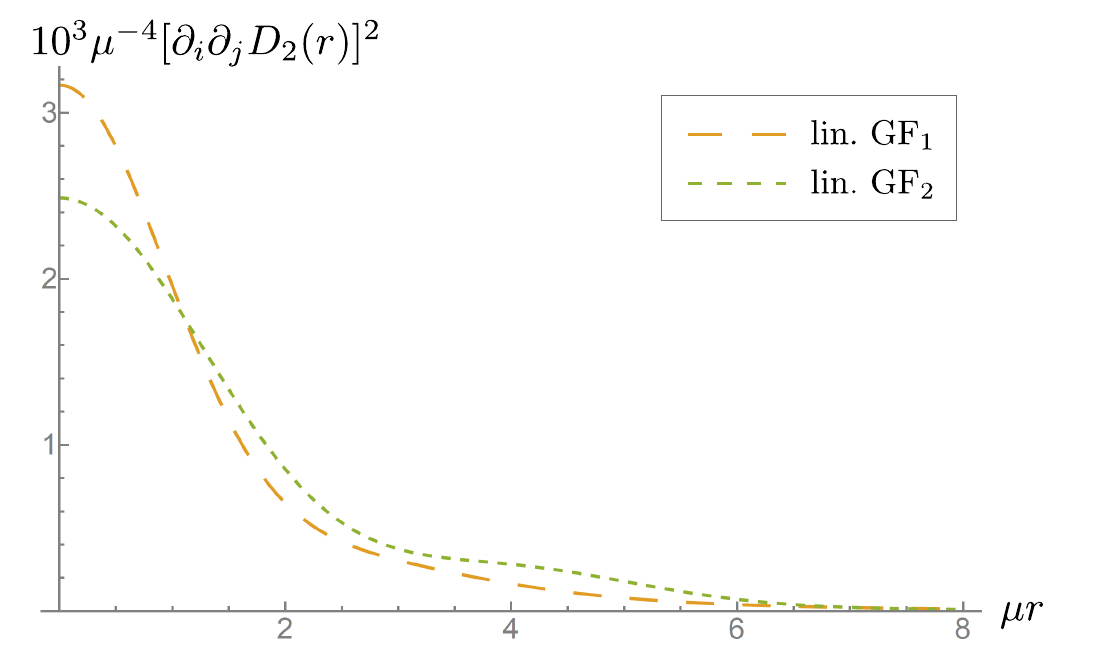} }}%
    \caption{Curvature invariants $I_2 = -\mu^{-2} \lap D_2(r)$ and $J_2 = \mu^{-4}[\partial_i \partial_j D_2(r)]^2$ for $\mathrm{GF_1}$ theory (dashed lines) and $\mathrm{GF_2}$ theory (dotted lines), which are relevant quantities for the cosmic string in $D=4$ as well as the angle deficit solutions in general $D$. The curvature invariants are smooth and well-behaved at $r=0$. Note that the invariants of type $I_m$ exhibit fluctuations to negative values at $\mu r \sim 5$, which might correspond to an ``anti-screening'' effect in GF gravity. For the plots of the other invariants, see appendix \ref{app:curvature-invariants-plots}.}
    \label{fig:visualization-curvature-invariants}
\end{figure*}

Substituting \eqref{eq:ansatz} and the interrelation of $u$ and $v$, see Eq.~\eqref{eq:u-v-relation}, into Eqs.~\eqref{eq:curv-riem}--\eqref{eq:curv-rscalar}, we find for the Ricci scalar as well as for the energy density
\begin{align}
\label{eq:energy-density}
R = - \lap v \, , \quad \rho_\ins{GR} = \frac{m+p-1}{2(1+p)} R \, .
\end{align}
The above justifies the interpretation of the Ricci scalar as a rescaled energy density. Moreover, the quadratic curvature invariants take the form
\begin{align}
C^2 &\equiv C_{\mu\nu\rho\sigma}C{}^{\mu\nu\rho\sigma} \nonumber \\
&= \frac{p^2-m^2+3m+p-2}{(1+p)(m+p)} \left( \lap v \right)^2 \\
&\hspace{11pt} + \frac{(m-2)(m+p-1)}{1+p} \left(\partial_i \partial_j v \right)\left( \partial^i \partial^j v \right) \, , \nonumber \\
\cancel{R}^2 &\equiv \cancel{R}{}_{\mu\nu}\cancel{R}{}^{\mu\nu} \nonumber \\
&= \frac{m(m+p-1)^2}{(1+p)(m+p+1)} \left( \lap v \right)^2 \, ,
\end{align}
where $\cancel{R}{}_{\mu\nu}$ denotes the tracefree Ricci tensor \eq{eq:tracefree-ricci}. In $D=4$, one may consider the Chern--Pontryagin pseudoscalar. It vanishes for warped geometries such as \eqref{eq:ansatz}:
\begin{align}
\mathcal{P} \equiv \frac12 \epsilon{}_{\mu\nu\alpha\beta} C{}^{\alpha\beta}{}_{\rho\sigma} C{}^{\mu\nu\rho\sigma} = 0 \, .
\end{align}
For more details of the curvature components, see appendix \ref{app:curvature}. Due to the radial dependence of the function $v = v(r)$, one has the following identities:
\begin{align}
\lap v &= v'' + (m-1) \frac{v'}{r} \, , \label{eq:derivative-simplification-1} \\
(\partial_i \partial_j v)(\partial^i \partial^j v) &= (v'')^2 + (m-1)\left(\frac{v'}{r}\right)^2 \, . \label{eq:derivative-simplification-2}
\end{align}
Using Eqs.~\eqref{eq:generate-green-functions} and \eqref{eq:general-result}, the above invariants can be rewritten in terms of higher-dimensional Green functions. Moreover, by virtue of Eq.~\eqref{eq:general-result} and the above relations, for a qualitative study of the behavior of the curvature invariants it is sufficient to substitute $D_m(r)$ into the expressions \eqref{eq:derivative-simplification-1} and \eqref{eq:derivative-simplification-2}. Hence let us consider the dimensionless invariants
\begin{align}
I^{(1,2)}_m &\equiv - \mu^{-m} \lap D^{(1,2)}_m(r) \, , \\
J^{(1,2)}_m &\equiv \mu^{-2m} \left( \partial_i \partial_j D^{(1,2)}_m(r) \right) \left( \partial^i \partial^j D^{1,2}_m(r) \right) \, .
\end{align}
We may think of the invariant $I^{(1,2)}_m$ as a rescaled energy density $\rho_\ind{GR}$ as per Eq.~\eqref{eq:energy-density}. Using Eq.~\eqref{eq:generate-green-functions} as well as Eqs.~\eqref{eq:derivative-simplification-1} and \eqref{eq:derivative-simplification-2}, the above invariants can be recast into the form
\begin{align}
I_m &= 2\pi \mu^{-m} \Big[ m D_{m+2}(r) - 2\pi r^2 D_{m+4}(r) \Big] \, , \\
J_m &= 4\pi^2 \mu^{-2m} \Big[ m D^2_{m+2}(r) - 4\pi r^2 D_{m+2}(r) D_{m+4}(r) \nonumber \\
&\hspace{105pt} + 4\pi^2 r^4 D^2_{m+4}(r) \Big] \, ,
\end{align}
where we suppressed the superscripts ``$(1,2)$'' for sake of clarity. In Fig.~\ref{fig:visualization-curvature-invariants} we visualize these invariants for some typical cases in $\mathrm{GF_1}$ and $\mathrm{GF_2}$ theory; for the other cases, see the plots in appendix \ref{app:curvature-invariants-plots}.

\subsection{Explicit examples}
Let us now discuss various examples of $p$-branes in $D$-dimensional Minkowski space in more detail. To that end, one may consider $p$-branes for different values of $p$, $m$ subject to $D = 1 + m + p$. In $D$ dimensions, we then have the following special cases of $p$-branes:
\begin{itemize}
\item point particle: ($p=0, m=D-1$),\\[-1.4\baselineskip]
\item cosmic string: ($p=1, m=D-2$),\\[-1.4\baselineskip]
\item domain wall: ($p=D-2, m=1$), \\[-1.4\baselineskip]
\item angle deficit configurations: ($p=D-3$, $m=2$).
\end{itemize}

\subsubsection{Point particle in any $D$}
A point particle can be regarded as the somewhat degenerate case of a 0-brane, and one can simply set $m=D-1$ such that $p=0$. This case has been previously studied in \cite{Frolov:2015usa}, and as follows from Eqs.~\eqref{eq:gf1-green-functions} and \eqref{eq:gf2-green-functions}, their gravitational potential is regularized. The Newtonian potential takes the form
\begin{align}
\Phi_\ind{N} = -\kappa\epsilon \frac{D-3}{D-2} D^{1,2}_{D-1}(r) \, ,
\end{align}
where $\epsilon$ has now dimensions of a mass. The above correctly reproduces the well-known trivial Newtonian limit in $D=3$ dimensions, as well as the modified Newtonian potential in, say, $\mathrm{GF_1}$ theory in $D=4$:
\begin{align}
\Phi_\ind{N} = -\text{erf}(\mu r/2) \frac{G\epsilon}{r} \rightarrow -\frac{G\epsilon}{r} \text{ for } \mu \rightarrow \infty \, .
\end{align}
The limiting case $\mu \rightarrow \infty$ reproduces the $1/r$ potential.

As was recently shown \cite{Buoninfante:2018xiw}, the Weyl tensor vanishes at the location of point-like particles in $\mathrm{GF_1}$ theory. Here, let us generalize this result in two directions by considering point particles in any dimension $D$ for a general $\mathrm{GF_N}$ theory. Assuming that the gravitational field of the point particle is regular \cite{Biswas:2011ar,Modesto:2010uh,Frolov:2015bia}, the metric function $v(r)$ has the following expansion at small distances:
\begin{align}
v(r) = a + b r^2  + \mathcal{O}\left(r^3\right) \, ,
\end{align}
where $a$ and $b$ are some constant factors. Then, to leading order, one has
\begin{align}
\partial_i \partial_j v = \frac 1m \delta_{ij} \lap v + \mathcal{O}\left( r \right) \, .
\end{align}
Upon this substitution it is easy to see that the Weyl tensor (appendix \ref{app:curvature}) vanishes identically iff $p=0$:
\begin{align}
C_{abcd} &= 0 \, , \\
C_{aibj} &= \frac{(m-1)p}{(m+p)(1+p)m} \eta{}_{ab} \delta{}_{ij} \lap v = 0 \, , \\
C_{ijkl} &= \frac{-p}{m(m+p)} 2 \delta_{i[k} \delta{}_{l]j} \lap v = 0 \, .
\end{align}
Therefore, as long as point particles have a regular gravitational field, the Weyl tensor vanishes at their location in $\mathrm{GF_N}$ theories. No such theorem holds for $p$-branes with $p \not= 0$, except when $m=1$, as we shall see below.

\subsubsection{Cosmic string in $D=4$}
A cosmic string in $D=4$, our example from the Introduction, is described by $p=1$ and $m=2$ such that
\begin{align}
v^{1,2}(r) = 2\kappa\epsilon D^{1,2}_2(r) \, , \quad u(r) = 0 \, .
\end{align}
The fact that $u$ vanishes corresponds to the absence of a gravitational field of the cosmic string \emph{in the Newtonian approximation} and in linearized GR. In GF theories, however, one may calculate that the curvature is non-zero, even for $u=0$. In the case of $\mathrm{GF_1}$ theory, the invariants take a simple form ($\zeta \equiv \mu r/2$):
\begin{align}
I^{(1)}_2 &= \frac{e^{-\zeta^2}}{4\pi} \, , \\
J^{(1)}_2 &= \frac{e^{-2\zeta^2}}{16\pi^2}\Big[ 1 + f(\zeta) + \frac 12 f(\zeta)^2 \Big] \, \\
f(\zeta) &\equiv \frac{1 - e^{\zeta^2}}{\zeta^2} \rightarrow -1 \text{ for } \zeta \rightarrow 0 \, .
\end{align}
Let us notice that for $m=2$ all expressions for the curvature invariants can be expressed solely in terms of $\lap v$, and $J_2$ does not enter them.

\subsubsection{Domain walls in any $D$, conformal flatness}
Domain walls are surfaces of dimension $D-1$ such that $p=D-2$ and $m=1$. Therefore one has
\begin{align}
v(r) = \frac{2\kappa\epsilon(D-1)}{D-2} D_1^{1,2}(r) \, , \quad u(r) = \frac{v(r)}{D-1} \, .
\end{align}
In this special case, the metric \eqref{eq:ansatz} is conformally flat. An easy way to see this is to introduce the new ``radial'' coordinate $y'$ according to
\begin{align}
\dd y \rightarrow \dd y' = \sqrt{\frac{1+u(y)}{1+v(y)}} \, \dd y \, ,
\end{align}
upon which the metric \eqref{eq:ansatz} becomes $g{}_{\mu\nu} = (1+u)\eta{}_{\mu\nu}$. Using the relations in appendix \ref{app:curvature}, as well as the one-dimensional identity $\partial_i \partial_j = \delta_{ij} \lap$, one can easily check that the Weyl tensor vanishes for $m=1$, as it must.

\subsubsection{Angle deficit configurations}
We have seen that a cosmic string, in the GR limit $\mu \rightarrow \infty$, does not gravitate but creates an angle deficit in the two-dimensional surface that it pierces. This triviality of the gravitational field is mirrored in the vanishing of the Newtonian potential for a cosmic string.

Lastly, let us generalize these solutions to the case $m=2$ for any $D$ such that $p=D-3$ and
\begin{align}
\Phi_\ind{N} = 0 \, .
\end{align}
The shape of the warped geometry \eqref{eq:ansatz} then guarantees that one still has an angle deficit of $\delta = -8\pi G$ in the $y_1y_2$-plane, regardless of the dimensionality of the $p$-brane.
However, as in the case of the cosmic string, the curvature invariants within GF theory do not vanish for this case.

\section{Conclusions}
\label{sec:conclusions}

In this paper we obtained and studied the gravitational field of infinitely thin objects such as point particles, strings, and branes, in the context linearized GF gravity. The characteristic property of these objects is that their stress-energy tensor is described by a $\delta$-function distribution localized at a point, straight line, or flat planes, respectively. We considered two versions of linearized GF gravity, $\mathrm{GF_1}$ and $\mathrm{GF_2}$, the form factors being
\begin{align}
a(\Box)=c(\Box) = \exp\left[(-\Box/\mu^2)^N\right] \, , \quad N=1,2 \, .
\end{align}
In the above, the parameter $\mu$ is the energy scale at which the GF theories start to deviate significantly from GR. Equivalently, one may think of $\ell = \mu^{-1}$ as the length scale of non-localities induced by the form factors $a(\Box)$ and $c(\Box)$. Consequently, for $\mu \rightarrow \infty$ one has $a=c\rightarrow 1$, thereby recovering GR in the infrared.

In our calculations we assumed that the thickness of the $p$-branes is much smaller than $\ell$, justifying the approximation of the stress-energy of the $p$-branes by simple $\delta$-distributions. In the linear approximation to the full GF gravity, the obtained Green functions can be used to generate solutions also in the case of thick $p$-branes.

The explicit $p$-brane solutions in linearized GF gravity in arbitrary number of spacetime dimensions $D$ possess two common properties: First, all of them are finite and regular at the position of the brane. Secondly, at far distances from the brane the GF gravity solutions asymptotically coincide with the solutions found in linearized GR (for the same, $\delta$-like stress-energy tensor). This second property is connected with the assumption that $a=c\rightarrow 1$ for $\mu \rightarrow \infty$, guaranteeing the proper infrared limit of the GF theory. The first property is a consequence of the adopted non-locality of the theory. It has a rather simple interpretation. The equation
\be
a(\lap)\lap=J
\ee
for a static source $J$ can be rewritten into the equivalent
\be
\lap=\tilde{J}\hh \tilde{J} \equiv a^{-1}(\lap)J\, .
\ee
Given a choice of non-local form factors, the effective current $\tilde{J}$ is smeared. This corresponds precisely to the quantity $\rho_\ind{GR}$ we defined in Sec.~\ref{sec:lin-curv}. As we have shown explicitly, this function is smooth for $\delta$-like sources. For point-like particles this effect is well known, and has been generalized in this work to static $p$-branes in $D$-dimensional Minkowski space.

Let us emphasize that the solutions for the gravitational potential of the $p$-branes are quite similar for $\mathrm{GF_1}$ and $\mathrm{GF_2}$ models. This happens because the sources are static. In the time-dependent case this is not true. In particular, this difference has been illustrated in case of the radiation of a time-dependent $\delta$-like source in the connection with a GF massless scalar field \cite{Frolov:2016xhq}. In general, $\mathrm{GF_1}$ theory is sensitive to the difference of the spacelike and time-like directions as a result of the Lorentz signature of the metric. $\mathrm{GF_2}$, on the other hand, has a better behavior. However, as argued above, for the static sources as considered in this work, this difference between $\mathrm{GF_1}$ and $\mathrm{GF_2}$ is irrelevant.

It might be interesting to study the gravitational field of $p$-branes in the complete, non-linear GF gravity. One might hope that such a problem can be treated because of the high symmetry of the $p$-brane sources. It is also interesting to consider linearized GF gravity solutions on constant curvature backgrounds like (A)dS and to study how GF gravity might manifest itself in the AdS/CFT correspondence.

\section*{Acknowledgments}

J.B.\ is grateful for a Vanier Canada Graduate Scholarship administered by the Natural Sciences and Engineering Research Council of Canada as well as for the Golden Bell Jar Graduate Scholarship in Physics by the University of Alberta.
V.F.\ and A.Z.\ thank the Natural Sciences and Engineering Research Council of Canada and the Killam Trust for their financial support.
V.F. also thanks the Yukawa Institute for Theoretical Physics at Kyoto University, where this work was completed during the workshop YITP-T-17-02 ``Gravity and Cosmology 2018.''
\vfill

\bibliography{Ghost_references}{}

\appendix

\section{Curvature expressions}
\label{app:curvature}
Let us calculate the curvature tensors for the metric
\begin{align}
\tag{\ref*{eq:ansatz}}
\begin{split}
\dd s^2 &= \hspace{11pt}(1 + u)\left( -\dd t^2 + \dd z_1^2 + \dots + \dd z_p^2 \right) \\
&\hspace{12pt} + (1 + v) \left( \dd y_1^2 + \dots + \dd y_m^2 \right) \,
\end{split}
\end{align}
subject to the substitution
\begin{align}
\tag{\ref*{eq:u-v-relation-simplified}}
u = \frac{2-m}{1+p} v \, .
\end{align}
We denote the flat background metric on the $tz_a$-sector as $\eta{}_{\bar{a}\bar{b}} = \text{diag}(-1,1,\dots,1)$, where now $\bar{a}=0,1,\dots,p$. The flat background metric on the $y_i$-sector is denoted by $\delta_{ij}$. Then, the Riemann tensor takes the form
\begin{align}
R_{\bar{a}\bar{b}\bar{c}\bar{d}} &= 0 \, , \\
R_{\bar{a}i\bar{b}j} &= \frac{m-2}{2(1+p)} \eta{}_{\bar{a}\bar{b}} \partial_i \partial_j v \, , \\
R{}_{ijkl} &= 2 \partial_{[i} \delta{}_{j][k} \partial_{l]} v \, .
\end{align}
The Ricci tensor is
\begin{align}
R_{\bar{a}\bar{b}} = \frac{m-2}{2(1+p)} \eta_{\bar{a}\bar{b}} \lap v \, , \quad R_{ij} = -\frac12 \delta_{ij} \lap v \, ,
\end{align}
and the Ricci scalar is simply
\begin{align}
R &= - \lap v \, .
\end{align}
We can construct the tracefree Ricci tensor:
\begin{align}
\cancel{R}_{\mu\nu} &\equiv R{}_{\mu\nu} - \frac1D R\, \eta_{\mu\nu} \, \label{eq:tracefree-ricci}\\
\cancel{R}_{\bar{a}\bar{b}} &= \frac{m(m+p-1)}{2(1+p)(m+p+1)} \eta_{\bar{a}\bar{b}} \lap v \, , \\
\cancel{R}_{ij} &= \frac{1-m-p}{2(m+p+1)} \delta_{ij} \lap v \, .
\end{align}
Let us define the Weyl tensor
\begin{alignat}{3}
C{}_{\mu\nu\rho\sigma} &= R{}_{\mu\nu\rho\sigma}&&- \frac{2}{D(D-1)} R\, \eta{}_{\mu[\rho}\eta{}_{\sigma]\nu} \\
& &&- \frac{2}{D-2} \left( \eta{}_{\mu[\rho} \cancel{R}_{\sigma]\nu} - \eta{}_{\nu[\rho} \cancel{R}_{\sigma]\mu} \right) , \nonumber \\
&= R{}_{\mu\nu\rho\sigma}&&+ \frac{2}{(D-1)(D-2)} R\, \eta{}_{\mu[\rho}\eta{}_{\sigma]\nu} \\
& &&- \frac{2}{D-2} \left( \eta{}_{\mu[\rho} R_{\sigma]\nu} - \eta{}_{\nu[\rho} R_{\sigma]\mu} \right) \, . \nonumber
\end{alignat}
For the components we obtain
\begin{align}
C_{\bar{a}\bar{b}\bar{c}\bar{d}} &= \frac{1-m}{(1+p)(m+p)} 2 \eta_{\bar{a}[\bar{c}} \eta_{\bar{d}]\bar{b}} \lap v \, , \\
C_{\bar{a}i\bar{b}j} &= \frac{m-2}{2(1+p)} \eta_{\bar{a}\bar{b}} \partial_i \partial_j v + \frac{2-m+p}{2(1+p)(m+p)} \eta_{\bar{a}\bar{b}} \delta_{ij} \lap v \, , \\
C_{ijkl} &= 2 \partial_{[i} \delta_{j][k} \partial_{l]} v + \frac{2}{m+p} \delta_{i[k} \delta_{l]j} \lap v \, .
\end{align}
Using the above relations, the expressions in Sec.~\ref{sec:lin-curv} can be readily derived.

\vfill

\pagebreak

\begin{widetext}
\section{Curvature invariants visualized}
\label{app:curvature-invariants-plots}

\begin{figure*}[!hbt]
    \centering
    \subfloat{{ \includegraphics[width=0.42\textwidth]{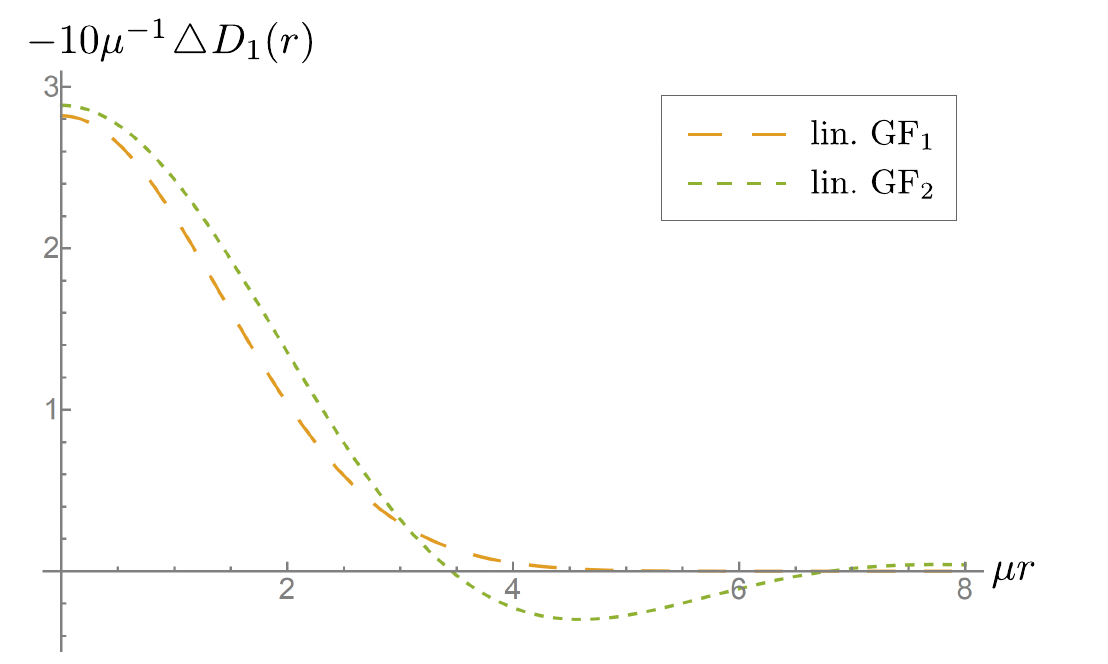} }}%
    \qquad
    \subfloat{{ \includegraphics[width=0.42\textwidth]{plot-i2.pdf} }}%

    \subfloat{{ \includegraphics[width=0.42\textwidth]{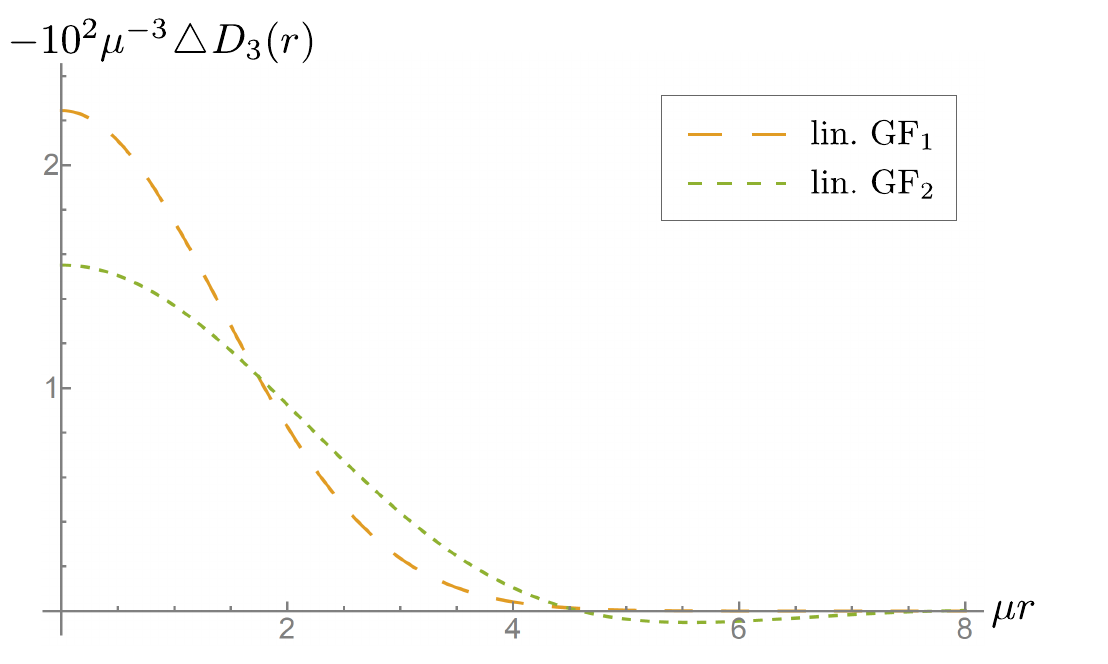} }}
    \qquad
    \subfloat{{ \includegraphics[width=0.42\textwidth]{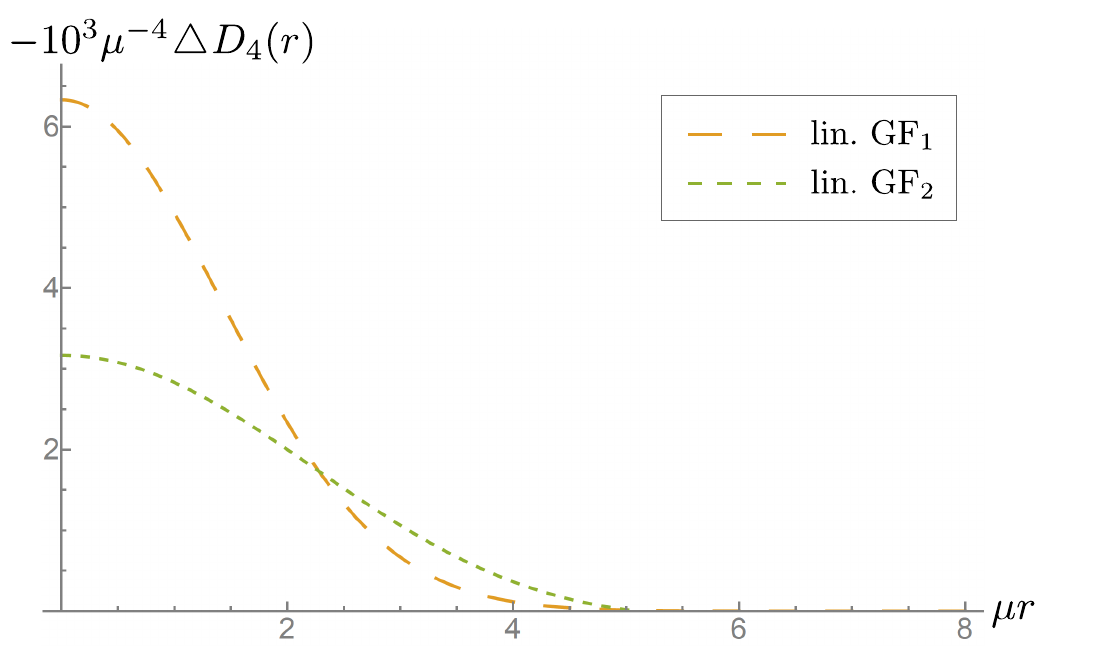} }}%
    \caption{Linear curvature invariant $I_m$ evaluated for $\mathrm{GF_1}$ and $\mathrm{GF_2}$ theory in the cases $m=1,2,3,4$.}
    \label{fig:visualization-i}
\end{figure*}

\begin{figure*}[!hbt]
    \centering
    \subfloat{{ \includegraphics[width=0.42\textwidth]{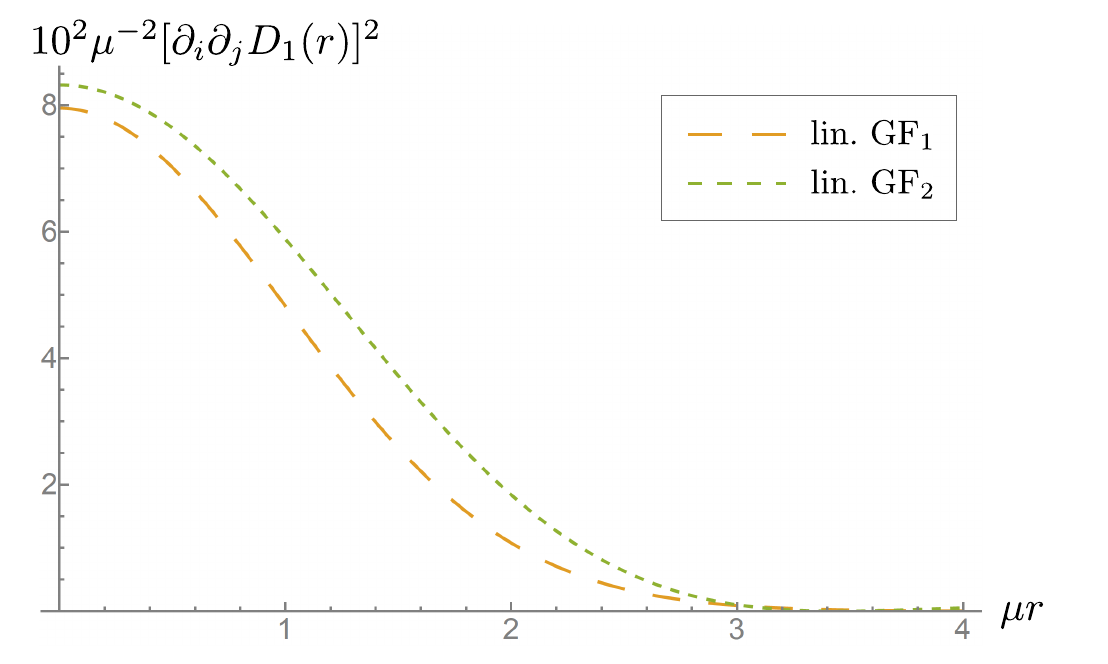} }}%
    \qquad
    \subfloat{{ \includegraphics[width=0.42\textwidth]{plot-j2.pdf} }}%

    \subfloat{{ \includegraphics[width=0.42\textwidth]{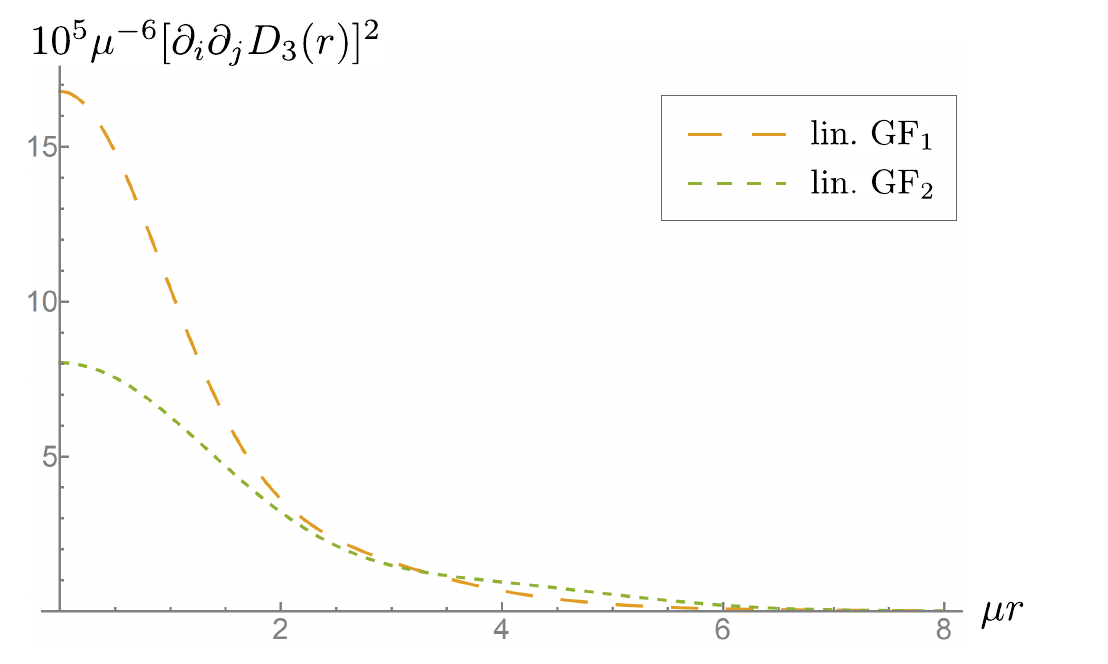} }}
    \qquad
    \subfloat{{ \includegraphics[width=0.42\textwidth]{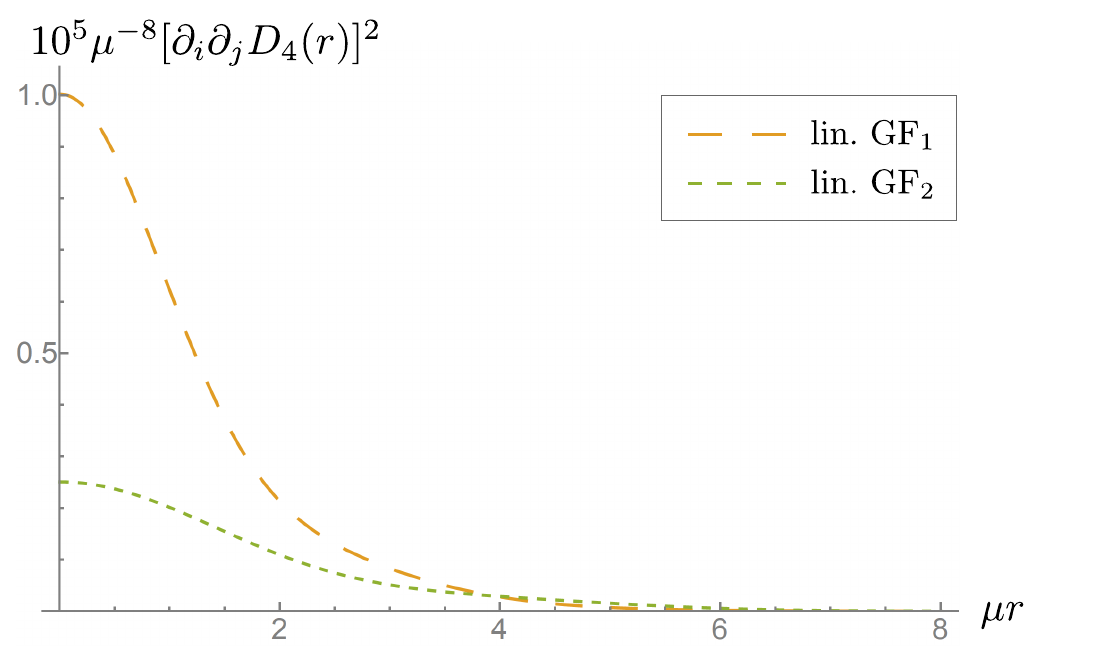} }}%
    \caption{Quadratic curvature invariant $J_m$ evaluated for $\mathrm{GF_1}$ and $\mathrm{GF_2}$ theory in the cases $m=1,2,3,4$.}
    \label{fig:visualization-j}
\end{figure*}
\end{widetext}

\end{document}